\documentclass[journal=jpclcd,manuscript=article]{achemso}
\setkeys{acs}{articletitle=true}

\usepackage[version=3]{mhchem} 

\usepackage{amsmath}
\usepackage{amsfonts}
\usepackage{amssymb}
\usepackage{graphicx}
\usepackage{dcolumn}
\usepackage{bm}
\usepackage{subcaption}
\usepackage{algorithm} 
\usepackage{algpseudocode} 
\usepackage{xr}
\usepackage{hyperref}
\usepackage{braket}
\usepackage{xcolor}

\makeatletter
\newcommand*{\addFileDependency}[1]{
  \typeout{(#1)}
  \@addtofilelist{#1}
  \IfFileExists{#1}{}{\typeout{No file #1.}}
}
\makeatother

\usepackage[capitalize]{cleveref}
\crefname{figure}{Fig.}{Figs.}
\Crefname{figure}{Figure}{Figures}
\crefname{table}{Tab.}{Tabs.}
\Crefname{table}{Table}{Tables}
\crefname{equation}{Eq.}{Eqs.}
\Crefname{equation}{Equation}{Equations}
\crefname{section}{Sec.}{Secs.}
\Crefname{section}{Section}{Sections}
\usepackage{xcolor}

\usepackage{array}
\newcommand{\PreserveBackslash}[1]{\let\temp=\\#1\let\\=\temp}
\newcolumntype{C}[1]{>{\PreserveBackslash\centering}p{#1}}
\newcolumntype{R}[1]{>{\PreserveBackslash\raggedleft}p{#1}}
\newcolumntype{L}[1]{>{\PreserveBackslash\raggedright}p{#1}}

\usepackage{multirow}

\author{Anton Nyk{\"a}nen}
\affiliation{\fontsize{10pt}{10pt}\selectfont Algorithmiq Ltd., Kanavakatu 3C, FI-00160 Helsinki, Finland}

\author{Aaron Miller}
\affiliation{\fontsize{10pt}{10pt}\selectfont Algorithmiq Ltd., Kanavakatu 3C, FI-00160 Helsinki, Finland}
\alsoaffiliation{\fontsize{10pt}{10pt}\selectfont School of Physics, Trinity College Dublin, College Green, Dublin 2, Ireland}

\author{Walter Talarico}
\affiliation{\fontsize{10pt}{10pt}\selectfont Algorithmiq Ltd., Kanavakatu 3C, FI-00160 Helsinki, Finland}
\alsoaffiliation{\fontsize{10pt}{10pt}\selectfont Department of Applied Physics, QTF Centre of Excellence, Center for Quantum
Engineering, Aalto University School of Science, FIN-00076 Aalto, Finland}

\author{Stefan Knecht}
\affiliation{\fontsize{10pt}{10pt}\selectfont Algorithmiq Ltd., Kanavakatu 3C, FI-00160 Helsinki, Finland}
\alsoaffiliation{\fontsize{10pt}{10pt}\selectfont ETH Z{\"u}rich, Department of Chemistry and Applied Life Sciences
Vladimir-Prelog-Weg 1-5/10, 8093 Z{\"u}rich, Switzerland}

\author{Arseny Kovyrshin}
\affiliation{\fontsize{10pt}{10pt}\selectfont Data Science and Modelling, Pharmaceutical Sciences, R\&D,
AstraZeneca Gothenburg, Pepparedsleden 1, Molndal SE-431 83, Sweden}

\author{Mårten Skogh}
\affiliation{\fontsize{10pt}{10pt}\selectfont Data Science and Modelling, Pharmaceutical Sciences, R\&D,
AstraZeneca Gothenburg, Pepparedsleden 1, Molndal SE-431 83, Sweden}
\alsoaffiliation{\fontsize{10pt}{10pt}\selectfont Department of Chemistry and Chemical Engineering, Chalmers University of Technology, Gothenburg, Sweden}

\author{Lars Tornberg}
\affiliation{\fontsize{10pt}{10pt}\selectfont Data Science and Modelling, Pharmaceutical Sciences, R\&D,
AstraZeneca Gothenburg, Pepparedsleden 1, Molndal SE-431 83, Sweden}

\author{Anders Broo}
\affiliation{\fontsize{10pt}{10pt}\selectfont Data Science and Modelling, Pharmaceutical Sciences, R\&D,
AstraZeneca Gothenburg, Pepparedsleden 1, Molndal SE-431 83, Sweden}

\author{Stefano Mensa}
\affiliation{\fontsize{10pt}{10pt}\selectfont The Hartree Centre, STFC, Sci-Tech Daresbury, Warrington, WA4 4AD, United Kingdom}

\author{Benjamin C. B. Symons}
\affiliation{\fontsize{10pt}{10pt}\selectfont The Hartree Centre, STFC, Sci-Tech Daresbury, Warrington, WA4 4AD, United Kingdom}

\author{Emre Sahin}
\affiliation{\fontsize{10pt}{10pt}\selectfont The Hartree Centre, STFC, Sci-Tech Daresbury, Warrington, WA4 4AD, United Kingdom}

\author{Jason Crain}
\affiliation{\fontsize{10pt}{10pt}\selectfont IBM Research Europe, Hartree Centre STFC Laboratory, Sci-Tech Daresbury, Warrington WA4 4AD, United Kingdom}
\alsoaffiliation{\fontsize{10pt}{10pt}\selectfont Department of Biochemistry, University of Oxford, Oxford, OX1 3QU, United Kingdom}

\author{Ivano Tavernelli}
\affiliation{\fontsize{10pt}{10pt}\selectfont IBM Quantum, IBM Research – Zürich, 8803 Rüschlikon, Switzerland}

\author{Fabijan Pavo\v{s}evi\'{c}}
\email{fpavosevic@gmail.com, fabijan.pavosevic@algorithmiq.fi}
\affiliation{\fontsize{10pt}{10pt}\selectfont Algorithmiq Ltd., Kanavakatu 3C, FI-00160 Helsinki, Finland}

\title[]
  {Toward Accurate Post-Born-Oppenheimer Molecular Simulations on Quantum Computers: An Adaptive Variational Eigensolver with Nuclear-Electronic Frozen Natural Orbitals}

\begin{document}



\begin{tocentry}
\begin{figure}[H]
	\begin{center}
		\includegraphics[width=1.7in]{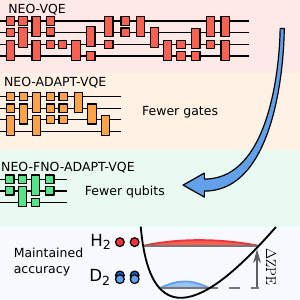}
	\end{center}
\end{figure}
\end{tocentry}

\begin{abstract}

Nuclear quantum effects such as zero-point energy and hydrogen tunnelling play a central role in many biological and chemical processes. The nuclear-electronic orbital (NEO) approach captures these effects by treating selected nuclei quantum mechanically on the same footing as electrons. On classical computers, the resources required for an exact solution of NEO-based models grow exponentially with system size. By contrast, quantum computers offer a means of solving this problem with polynomial scaling. However, due to the limitations of current quantum devices, NEO simulations are confined to the smallest systems described by minimal basis sets whereas realistic simulations beyond the Born--Oppenheimer approximation require more sophisticated basis sets. For this purpose, we herein extend a hardware-efficient ADAPT-VQE method to the NEO framework in the frozen natural orbital (FNO) basis. We demonstrate on H$_2$ and D$_2$ molecules that the NEO-FNO-ADAPT-VQE method reduces the CNOT count by a several orders of magnitude relative to the NEO unitary coupled cluster method with singles and doubles (NEO-UCCSD) while maintaining the desired accuracy. This extreme reduction in the CNOT gate count is sufficient to permit practical computations employing the NEO method --- an important step toward accurate simulations involving non-classical nuclei and non--Born--Oppenheimer effects on near-term quantum devices. We further show that the method can capture isotope effects and we demonstrate that inclusion of correlation energy systematically improves the prediction of $\Delta$ZPE.  

\end{abstract}

\maketitle

\section{Introduction}

Most quantum chemistry simulations invoke the Born--Oppenheimer approximation which assumes that electrons respond instantaneously to a change in position of the nuclei. By contrast, a wide range of important chemical and biological phenomena, such as proton-coupled electron transfer or proton tunneling, require a quantum mechanical treatment of both electrons and nuclei that goes beyond the Born--Oppenheimer approximation.~\cite{cha1989hydrogen,tuckerman1997quantum,hammes2010theory,weinberg2012proton} This has provided a sustained impetus for development of accurate theoretical methods to simulate such processes.~\cite{ishimoto2009review,abedi2010exact,yonehara2012fundamental,habershon2013ring,curchod2018ab,pavosevic2020chemrev,reyes2019any,ollitrault2022quantum} Among different methods, the nuclear-electronic orbital (NEO)~\cite{webb2002multiconfigurational,pavosevic2020chemrev} framework treats specified nuclei quantum mechanically on an equal footing with electrons using molecular orbital (MO) techniques, while avoiding the Born--Oppenheimer separation between their degrees of freedom. It, therefore, offers a compromise between computational efficiency and accuracy for incorporation of nuclear quantum effects, such as zero-point energy (ZPE), proton delocalization, vibrational anharmonicity, and isotope effects.~\cite{pavosevic2020chemrev} Both wave function-based methods~\cite{webb2002multiconfigurational,nakai2003many,pavosevic2018ccsd,pavovsevic2019neoeomccsd,pavovsevic2019neobccd,pavosevic2020neooomp2,pavosevic2021neodfccsd,pavosevic2020neortccsd,pavovsevic2020automatic,fajen2020separation,muolo2020nuclear,fajen2021multicomponent,alaal2021multicomponent,fetherolf2022multicomponent,pavovsevic2022neoccsdt,fowler2022t,feldmann2022quantum,feld23a} and density functional theory (DFT)~\cite{pak2007density,yang2017development,brorsen2017multicomponent,yang2018multicomponent} have been implemented within the NEO framework, allowing for an accurate description of nuclear quantum effects of molecules in the ground and excited states. An advantage of wave function-based NEO methods over their NEO-DFT counterparts is that they are systematically improvable; which ultimately leads to the exact solution of coupled nuclear-electron molecular systems. 

As in conventional electronic structure theory, the exact solution of wave function-based NEO methods exhibits exponential scaling with system size on classical computers. 
However, quantum computers, in conjunction with the quantum phase estimation (QPE) algorithm,~\cite{aspuru2005simulated} offer a means of solving this problem at a cost that scales polynomially with system size.~\cite{veis2016quantum} Unfortunately, implementing QPE on quantum hardware requires logical qubits and deep circuits~\cite{bauer2020quantum}, at resource levels that exceed the capabilities of currently available near-term quantum devices.~\cite{preskill2018quantum} By contrast, the hybrid variational quantum eigensolver (VQE)~\cite{peruzzo2014variational} (utilizing a quantum device for wave function preparation and ground-state energy estimation, and a classical computer for the variational optimization of wave function parameters) is well suited for current noisy quantum devices due to relatively low circuit depths. 

To enable a near-exact simulation of molecular systems beyond the Born--Oppenheimer approximation on present-day quantum devices, one of us (FP) has recently developed the NEO unitary coupled cluster method within the VQE algorithm (NEO-UCC/VQE).~\cite{pavosevic2021neouccsd} The developed NEO-UCC/VQE method was applied to study positronium hydride as well as H$_2$\ where, in addition to two electrons, one positron or one proton were treated quantum mechanically, respectively.~\cite{pavosevic2021neouccsd} This study demonstrated that the NEO-UCC/VQE ground and excited state energies for these two systems are in excellent agreement with the exact (full CI) energies.~\cite{pavosevic2021neouccsd} Following that work, qubit tapering and ansatz parameter initialization procedures were introduced,~\cite{kovyrshin2023quantum} eventually leading to the development of an approach~\cite{Kovyrshin2023jpcl} to simulate proton transfer dynamics. The first impediment of the discussed simulations is that they are limited to minimal or near-minimal electronic and nuclear basis sets. Although the use of minimal basis sets is useful in proof of concept demonstrations, in order to achieve quantitatively accurate results comparable to experimentally measured quantities, one must employ much larger and far more flexible electronic and nuclear basis sets.~\cite{pavovsevic2022neoccsdt,fowler2022t,samsonova2023hydrogen} The second impediment is that the VQE algorithm employs the NEO-UCC ansatz: In spite of the fact that NEO-UCC/VQE exhibits high accuracy compared to the exact solution,~\cite{pavosevic2021neouccsd} practical implementations of the NEO-UCC/VQE method on current quantum hardware would be limited only to the smallest systems due to  the prohibitively deep quantum circuits involved.~\cite{kovyrshin2023quantum} 

To facilitate quantitatively accurate molecular simulations beyond the Born--Oppenheimer approximation on noisy, near-term quantum devices, herein we address the two aforementioned impediments. To overcome the first roadblock and to make simulations beyond minimal basis sets feasible, we introduce NEO frozen natural orbitals (FNO). These are designed to enable effective truncation of the unoccupied electronic and nuclear orbital spaces. The FNO approximation thereby offers an efficient and robust means of compressing the number of unoccupied orbitals,~\cite{sosa1989selection} and has become a popular choice for enhancing the computational efficiency of electronic structure methods for classical~\cite{taube2005frozen,landau2010frozen,deprince2013accurate} and quantum computers.~\cite{verma2021scaling,metcalf2020resource} To address the second issue, we propose an implementation of the Adaptive Derivative Assembled Problem-Tailored Ansatz Variational Quantum Eigensolver~\cite{grimsley2019adaptive,tang2021qubit} algorithm within the NEO framework (referred to as NEO-ADAPT-VQE). 
Instead of relying on deep fixed circuits from the NEO-UCC ansatz, NEO-ADAPT-VQE iteratively adds the excitation operators with the highest energy gradient to the ansatz until convergence. Therefore, it significantly reduces the quantum hardware requirements as the final circuit is usually considerably shallower~\cite{grimsley2019adaptive} than in the case of the NEO-UCC/VQE method.

\begin{figure*}[ht!]
  \centering
  \includegraphics[width=6.5in]{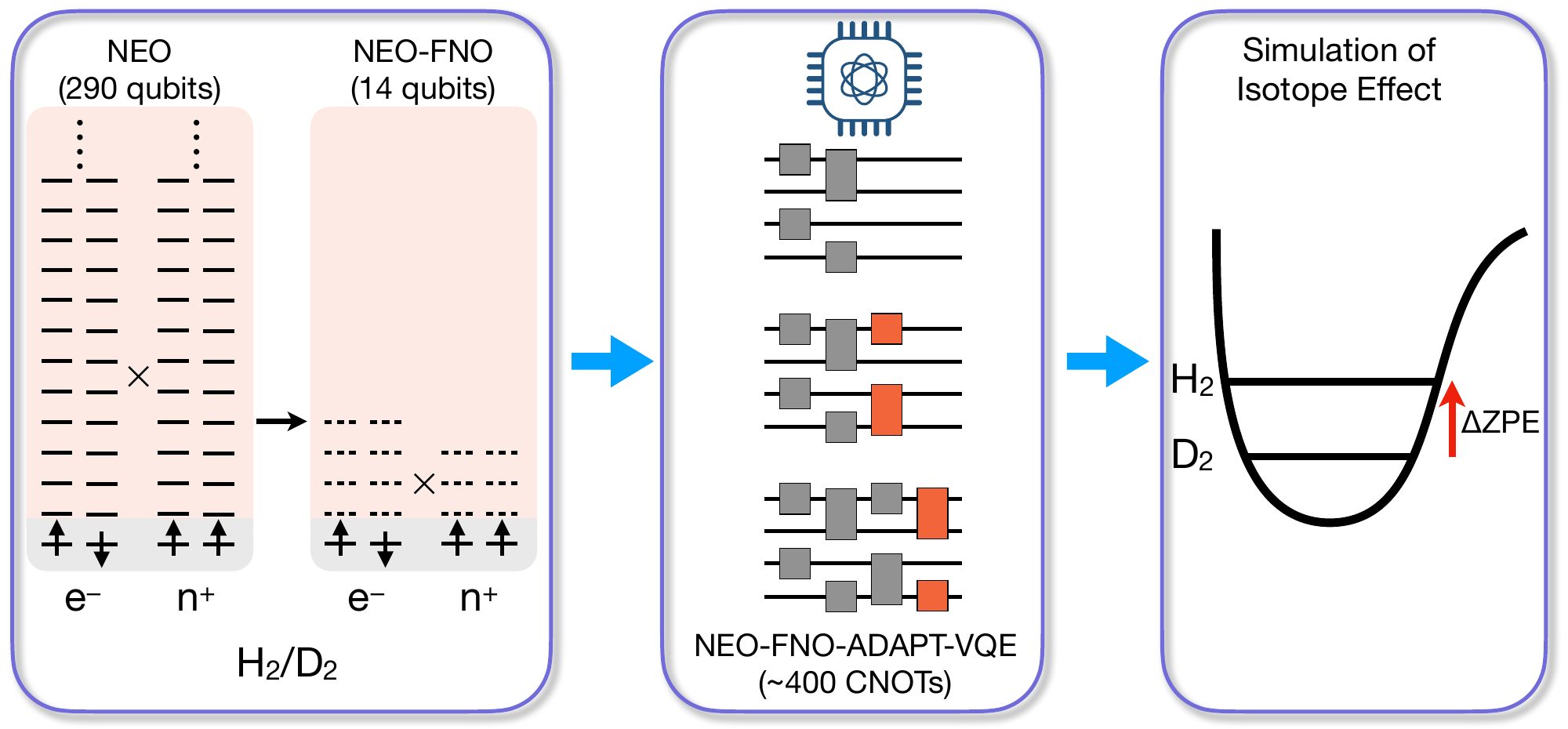}
  \caption{Schematic representation of the steps performed in this work. The left panel illustrates a large orbital NEO space followed by a compressed representation in the FNO basis which is then used to perform the NEO-FNO-ADAPT-VQE simulation (middle panel). In the final step, the energy difference between the two isotopes is calculated.}
  \label{fig:schematic}
\end{figure*}

To demonstrate the performance and accuracy of the developed NEO-ADAPT-VQE method within the FNO approximation, in this work we study the isotope effect for H$_2$ and D$_2$. To this end, Figure~\ref{fig:schematic} displays a schematic representation of the NEO-FNO-ADAPT workflow. The developments and detailed analysis described in this work highlight the robustness and reliability of the proposed method for simulation of nuclear quantum effects on contemporary quantum computers. Moreover, this work paves the way for a wide range of developments and applications on classical and quantum computers within the NEO framework. 

\section{Theory}

In the VQE algorithm, the NEO ground state energy $E_{\text{NEO-VQE}}$ is determined via the variational optimization of the following energy functional
\begin{equation}
    \label{eqn:NEO-VQE-Energy-Funtional}
     E_{\text{NEO-VQE}}=\underset{\theta}{\text{min}} \langle\Psi_{\text{NEO}}(\theta)|\hat{H}_{\text{NEO}}|\Psi_{\text{NEO}}(\theta)\rangle
\end{equation}
with respect to the wave function parameters $\theta$. Here, $\hat{H}_{\text{NEO}}$ is the NEO Hamiltonian for two different types of particles (i.e. electrons and protons, but extension to other types of fermionic particles is straightforward~\cite{pavosevic2020chemrev}) treated quantum mechanically. Within the second-quantized formalism, the NEO Hamiltonian is expressed as
\begin{equation}
    \label{eqn:NEO-Hamiltonian}
    \hat{H}_\text{NEO} = h^p_q a^q_p + \frac{1}{2}g^{pq}_{rs}a^{rs}_{pq}+h^P_Q a^Q_P + \frac{1}{2}g^{PQ}_{RS}a^{RS}_{PQ} - g^{pP}_{qQ} a^{qQ}_{pP}
\end{equation}
where $a_{p_1p_2...p_n}^{q_1q_2...q_n}=a_{q_1}^{\dagger}a_{q_2}^{\dagger}...a_{q_n}^{\dagger}a_{p_n}...a_{p_2}a_{p_1}$ are second-quantized electronic excitation operators defined in terms of fermionic creation ($a^{\dagger}$) and annihilation ($a$) operators. The corresponding protonic excitation operators are defined analogously. In Eq.~\eqref{eqn:NEO-Hamiltonian}, $h^p_q=\langle q |\hat{h}^\text{e}|p \rangle$ and $h^P_Q=\langle Q |\hat{h}^\text{p}|P \rangle$ are electronic and protonic core Hamiltonian matrix elements, respectively. Additionally, $g^{pq}_{rs}=\langle rs|pq \rangle$, $g^{PQ}_{RS}=\langle RS|PQ \rangle$, and $g^{pP}_{qQ}=\langle qQ|pP \rangle$ are the electron-electron repulsion, the proton-proton repulsion, and the electron-proton attraction tensor elements, respectively. In this work,
the lower-case indices $p,q,r,s,...$, $i,j,k,l,...$, and $a,b,c,d,...$ denote general, occupied, and unoccupied (virtual) electronic spin orbitals, respectively, whereas the upper-case indices denote the protonic spin orbitals and are defined analogously. Throughout this paper, the Einstein summation over repeated indices is assumed. 

In Eq.~\eqref{eqn:NEO-VQE-Energy-Funtional}, $|\Psi_{\text{NEO}}(\theta)\rangle$ is the trial wave function, that in case of the NEO-UCC method takes the following form:
\begin{equation}
    \label{eqn:NEO-UCC}
    |\Psi_{\text{NEO-UCC}}\rangle=e^{\hat{T}-\hat{T}^\dagger}|0^{\text{e}}0^{\text{p}}\rangle
\end{equation}
In this equation, $\hat{T}=\theta_{\mu}a^{\mu}$ is the excitation cluster operator that incorporates the correlation effects between quantum particles (i.e. electrons and protons) by acting on the reference NEO-Hartree-Fock ($|0^{\text{e}}0^{\text{p}}\rangle=|0^{\text{e}}\rangle\otimes|0^{\text{p}}\rangle$) state composed from the electronic ($|0^{\text{e}}\rangle$) and protonic ($|0^{\text{p}}\rangle$) Slater determinants. Moreover, $\theta_{\mu}$ are the unknown wave function parameters determined from Eq.~\eqref{eqn:NEO-VQE-Energy-Funtional} and $a^{\mu}=a_{\mu}^{\dagger}\in \{a_{i}^{a},a_{I}^{A},a_{ij}^{ab},a_{IJ}^{AB},a_{iI}^{aA},...\}$ is a set of single (electron, proton), double (electron-electron, proton-proton, electron-proton), and higher excitation operators, whereas $\mu$ is an excitation manifold. Retaining only single and double excitations in the cluster expansion, defines the NEO unitary coupled cluster with singles and doubles (NEO-UCCSD) method.~\cite{pavosevic2021neouccsd,kovyrshin2023quantum} Due to the unitary nature of the NEO-UCCSD ansatz, the method is in principle suitable for quantum computations. However, practical implementations of the NEO-UCCSD/VQE method are limited only to the smallest systems in minimal electronic and nuclear basis sets due to a large number of wave function parameters and corresponding deep circuits.~\cite{kovyrshin2023quantum} 

To resolve these shortcomings, we extend the hardware-efficient ADAPT-VQE method~\cite{grimsley2019adaptive,tang2021qubit} to within the NEO framework. Analogous to its electronic counterpart~\cite{grimsley2019adaptive}, in the NEO-ADAPT-VQE algorithm, the NEO wave function ansatz grows systematically, and after the $n$-th iteration, the NEO-ADAPT ansatz is given by
\begin{equation}
    \label{eqn:NEO-ADPT}
    |\Psi_{\text{NEO-ADAPT}}^{(n)}\rangle=e^{\theta_n\tau^n}...e^{\theta_2\tau^2}e^{\theta_1\tau^1}|0^{\text{e}}0^{\text{p}}\rangle
\end{equation}
At every iteration, the fermionic operator that contributes the most toward lowering the NEO-VQE energy is selected from the fermionic operator pool $\tau^\mu=-\tau_\mu^{\dagger}\in\{a_{i}^{a}-a^{i}_{a},a_{I}^{A}-a^{I}_{A},a_{ij}^{ab}-a^{ij}_{ab},a_{IJ}^{AB}-a^{IJ}_{AB},a_{iI}^{aA}-a^{iI}_{aA}\}$ and added to the ansatz. The change in energy for a given operator is estimated by evaluating the energy gradient with respect to the corresponding $\mu$-th variational parameter as
\begin{equation}
    \label{eqn:NEO-ADPT-Grad}
    \frac{\partial E^{(n)}}{\partial \theta_{\mu}} = \langle\Psi^{(n)}_{\text{NEO-ADAPT}}| [\hat{H}_{\text{NEO}},\tau^{\mu}] |\Psi^{(n)}_{\text{NEO-ADAPT}}\rangle
\end{equation}
The ansatz grows until the norm of the energy gradient becomes smaller than a predefined threshold or until other selection criteria are satisfied (see below for more explanations).
By means of fermion-to-qubit mapping~\cite{Jordan1928, bonsai}, the excitation operators are converted to the qubit basis, $\tau^{\mu} \mapsto \sum_k c_k S_k$, where each $S_k$ represents a string composed of Pauli matrices $I, X, Y, Z$. Within this qubit space, we can further break down the qubit operator pool as $\tau = i\prod_{m} p_m$~\cite{ryabinkin2018qubit,tang2021qubit}, where each $p_m$ belongs to the set of Pauli matrices ${I, X, Y, Z}$. 
Utilizing a qubit pool of operators generally simplifies the gate complexity compared to a fermionic pool but increases the number of variational parameters in the ansatz. In what follows, we will refer to f-NEO-ADAPT when the operators are selected from a fermionic pool in contrast to q-NEO-ADAPT which denotes a selection of operators from a qubit pool. In both cases, the key feature of the NEO-ADAPT-VQE method is the low number of variational wave function parameters and reduced circuit depth compared to NEO-UCCSD, thereby rendering the former suitable for efficient molecular simulations beyond the Born--Oppenheimer approximation on contemporary quantum devices.

In the NEO framework, the description of the system consists of $\alpha$- and $\beta$-electrons as well as $\alpha$-protons.~\cite{pavosevic2020chemrev} From the three particle types we can recognize three $\mathbb{Z}_2$-symmetries for the particle number operators in the three subspaces.~\cite{kovyrshin2023quantum} All present $\mathbb{Z}_2$-symmetries of the system can be found by applying the procedure introduced by Bravyi et al.~\cite{bravyi2017tapering} to the Hamiltonian. This applies a Clifford transformation to the Hamiltonian which allows some of the qubits to be treated classically and tapered off. Once the tapering procedure has been performed on the Hamiltonian, the same procedure needs to be performed on all operators in the operator pool of NEO-ADAPT-VQE. Further qubits may be tapered by exploiting molecular point group symmetries within the NEO framework as discussed in Ref.~\citenum{kovyrshin2023quantum}.

To achieve high accuracy in quantum chemistry calculations and to recover a large portion of the dynamical correlation energy, it is necessary to employ extensive and flexible basis sets. These are spanned by a large number of unoccupied orbitals thereby increasing the number of qubits required for a quantum simulation of the full system. As already discussed in the introduction, the frozen natural orbital (FNO) approximation offers a robust way for reducing the number of unoccupied orbitals.~\cite{sosa1989selection} Herein, we extend the FNO approximation within the NEO framework in which two different sets of FNOs are constructed --- one corresponding to the electronic unoccupied orbitals, and the other to the protonic unoccupied orbitals. The electronic and protonic FNOs are then defined as the eigenvectors of the virtual-virtual block of the electronic one-particle density matrix
\begin{equation}
    \label{eqn:electronic-1rdm}
    \gamma_a^b=\langle\Psi_{\text{NEO}}|a_a^b|\Psi_{\text{NEO}}\rangle
\end{equation}
and the protonic one-particle density matrix
\begin{equation}
    \label{eqn:protonic-1rdm}
    \gamma_A^B=\langle\Psi_{\text{NEO}}|a_A^B|\Psi_{\text{NEO}}\rangle\
\end{equation}
respectively, and the eigenvalues correspond to the occupation numbers. Because the occupation numbers are proportional to the wave function expansion coefficients, the corresponding eigenvalues are used as criteria for truncation of the FNOs. Therefore, the FNOs for which the occupation number is smaller than the predefined threshold are neglected without introducing a significant error. In this work, the density matrices in Eq.~\eqref{eqn:electronic-1rdm} and Eq.~\eqref{eqn:protonic-1rdm} are computed with a low-cost NEO scaled-opposite-spin first-order Møller--Plesset (NEO-SOS-MP2)~\cite{pavosevic2020neooomp2,fetherolf2022multicomponent} wave function ($|\Psi_{\text{NEO}}\rangle$), and their programmable expressions are provided in Eqs.~S56 and~S58 of Ref.~\citenum{pavosevic2020neooomp2}, respectively. The NEO-SOS-MP2 method can be implemented with $\mathcal{O}(N^4)$ scaling, where $N$ is a measure of the system size. The electron-nuclear scaling parameters of the NEO-SOS-MP2 method~\cite{fetherolf2022multicomponent} employed in this work are 2.0 and 1.6 for proton and deuterium, respectively. Because the NEO-ADAPT-VQE correlation energy is calculated using a reduced number of FNOs, the full correlation energy is calculated by adding the correction, $\Delta E_{\text{NEO-SOS-MP2}}=E_{\text{NEO-SOS-MP2}}^{\text{MO}}-E_{\text{NEO-SOS-MP2}}^{\text{FNO}}$, that is defined as the difference between a nontruncated (in the MO basis, $E_{\text{NEO-SOS-MP2}}^{\text{MO}}$) and truncated (in the FNO basis, $E_{\text{NEO-SOS-MP2}}^{\text{FNO}}$) correlation energy~\cite{neese2009efficient}. All of the reported energy values in the FNO basis presented in the main paper include this correction, whereas energies without this correction are provided in the Supporting Information.

\section{Results and Discussion}
The NEO-ADAPT-VQE method was implemented within Algorithmiq’s software framework \textsf{Aurora}. Throughout this work, we employ a noise-free quantum statevector simulator. All of the calculations were performed on H$_2$/D$_2$ molecules where both the electronic and nuclear basis sets were centered at the hydrogen/deuterium positions with an inter-atomic distance of 0.7414~\AA. The reported calculations employ the cc-pV5Z~\cite{dunning1989gaussian} electronic basis set and the PB4-F2 (4s3p2d2f) nuclear basis set~\cite{yu2020development}. 
The one-particle ($h$) and two-particle ($g$) molecular integrals in Eq.~\eqref{eqn:NEO-Hamiltonian} were obtained from NEO-Hartree-Fock (NEO-HF) calculations using the \textsf{Q-Chem} quantum chemistry software~\cite{epifanovsky2021software}. While it is possible to eliminate global translational and rotational contributions from the NEO Hamiltonian (which spuriously increase energy), as indicated in Ref.~\citenum{naka2005}, this falls beyond the scope of the present study. The fermionic operators of the Hamiltonian and the excitation operator pool were mapped into the qubit space using the Parity transformation~\cite{bravyi2017tapering}. The wave function parameter optimizations for the NEO-UCCSD and NEO-ADAPT-VQE methods were performed with the L-BFGS-B optimizer~\cite{Zhu1997}, while the gradient calculations took advantage of the analytic approach introduced in Ref.~\citenum{grimsley2019adaptive}. The CNOT-counts were estimated by first compiling the circuits with \textsf{TKET} compilation pass outlined in Ref.~\citenum{Cowtan2020AGC} followed by the \textsf{Qiskit} transpile pass~\cite{Qiskit}.

First, we discuss the effect of the FNO approximation on the calculation of energy. The exact reference energies for the H$_2$ and D$_2$ molecules were calculated with the projective NEO Full Configuration Interaction (NEO-FCI) method implemented in \textsf{Python}. The programmable NEO-FCI equations have been derived with the \textsf{SeQuant} package~\cite{githubsequant}, and the details of the projective NEO-FCI implementation are described in Ref.~\citenum{pavosevic2018ccsd}. The NEO-FCI energies with cc-pV5Z/PB4-F2 basis sets are provided in Table~\ref{table:table1}. For comparison, we also include the NEO-Hartree-Fock (NEO-HF), NEO configuration interaction with singles and doubles (NEO-CISD), and NEO coupled cluster with singles and doubles (NEO-CCSD) energies~\cite{pavosevic2018ccsd}. Given such a basis set pair, the number of electronic and nuclear spin orbitals is 220 and 74, respectively. Therefore, a simulation on a quantum device would require 290 qubits with the tapering procedure introduced in the previous section. To arrive at a manageable qubit count, we reduce the number of electronic and nuclear unoccupied orbitals by truncating all of the FNOs with occupation numbers below $1\mathrm{e}{-3}$ and $1\mathrm{e}{-4}$, respectively. As a result, the orbital space in the FNO basis reduces substantially to merely 10 electronic and 8 nuclear spin orbitals. Within this set of FNO orbitals, a NEO-FNO-FCI calculation yields approximately 90\% of the total NEO-FCI correlation energy and the error relative to NEO-FCI for both H$_2$\ and D$_2$\ is less than 10~mHartree after the $\Delta E_{\text{NEO-SOS-MP2}}$ correction is included (see Table~\ref{table:table1}). Table~S1 shows that without $\Delta E_{\text{NEO-SOS-MP2}}$ around 60\% of the NEO-FCI correlation energy is recovered. For comparison, Table~S1 also shows that the nuclear electronic complete active space configuration interaction (NEO-CASCI) method comprised of the same number of active NEO-HF orbitals recovers only around 10\% of the total NEO-FCI correlation energy, therefore unequivocally demonstrating the benefit of using the FNO orbitals. Lastly, an increase in the number of FNOs would systematically reduce the discrepancy of the NEO-FNO-FCI method relative to NEO-FCI.  

\begin{table}[!htb]

\caption{Ground state energies (in Hartree) and its differences ($\Delta$ZPE in kcal/mol) for H$_2$ and D$_2$ molecules calculated with the NEO-HF, NEO-CISD, NEO-CCSD, NEO-FCI, NEO-FNO-FCI,\textsuperscript{a} NEO-FNO-UCCSD,\textsuperscript{a} f-NEO-FNO-ADAPT,\textsuperscript{a} and q-NEO-FNO-ADAPT\textsuperscript{a} methods employing the cc-pV5Z electronic and PB4-F2 nuclear basis sets. Required CNOT counts are given in parentheses.}

\begin{tabular}{C{4.2cm}|C{3.2cm}|C{3.2cm}|C{2cm}}
\hline
 & H$_2$ & D$_2$ & $\Delta\text{ZPE}$\\ \hline\hline
NEO-HF & -1.0519471 & -1.0738798 & 13.8\\ \hline
NEO-CISD & -1.1305647 & -1.1412573 & 6.7\\ \hline
NEO-CCSD & -1.1418725 & -1.1501648 & 5.3\\ \hline
NEO-FCI &  -1.1480488 & -1.1543290 & 3.9\\ \hline
NEO-FNO-FCI & -1.1385032 & -1.1451751 & 4.2\\ \hline
NEO-FNO-UCCSD & -1.1382860 (4938) & -1.1450513 (4940) & 4.2\\ \hline
f-NEO-FNO-ADAPT\textsuperscript{b} & -1.1369838 (872) & -1.1436931 (788) & 4.2\\ \hline
q-NEO-FNO-ADAPT\textsuperscript{b} & -1.1369342 (420) & -1.1436187 (410) & 4.2\\ \hline
Theory & -1.1640250~\cite{bubin2009non} & -1.1671688~\cite{bubin2010accurate}& 2.0\\ \hline\hline
\end{tabular}

\raggedright \textsuperscript{a}\small The orbital space in the FNO basis is comprised of 10 electronic and 8 nuclear spin orbitals. The reported energy values in the FNO basis include the $\Delta E_{\text{NEO-SOS-MP2}}$ correction.\\
\raggedright \textsuperscript{b}\small The reported f-NEO-FNO-ADAPT and q-NEO-FNO-ADAPT energy values are for which the energy difference relative to NEO-FNO-FCI is below 1.6~mHartree.\\

\label{table:table1}
\end{table}

We now focus attention on the NEO-FNO-ADAPT-VQE quantum simulation. The truncated fermionic NEO Hamiltonian in the FNO basis was mapped into the qubit space using the parity transformation~\cite{bravyi2017tapering} leading to a total of 18 qubits. Applying the tapering procedure introduced in the previous section, four $\mathbb{Z}_2$-symmetries were found, thus reducing the number of qubits to 14. An equivalent mapping and tapering procedure was carried out for the construction of the fermionic operator pool that contains the excitation operators. In the NEO-FNO-ADAPT-VQE simulations, the ansatz was first initialised to represent the NEO-HF state, while each ensuing iteration consisted of the following steps: 1) Measure the current energy of the system; 2) Estimate gradients of all operators in the pool by measuring their commutators with the Hamiltonian; 3) If the gradient norm of the operator pool is below a certain threshold, abort the algorithm; 4) Pick the operator with the highest gradient and add it to the ansatz with parameter 0; 5) Optimise all parameters in the ansatz. In addition, the simulations were aborted when the energy reached the known exact energy up till a desired numerical precision. 

\begin{figure*}[ht!]
  \centering
  \includegraphics[width=6.5in]{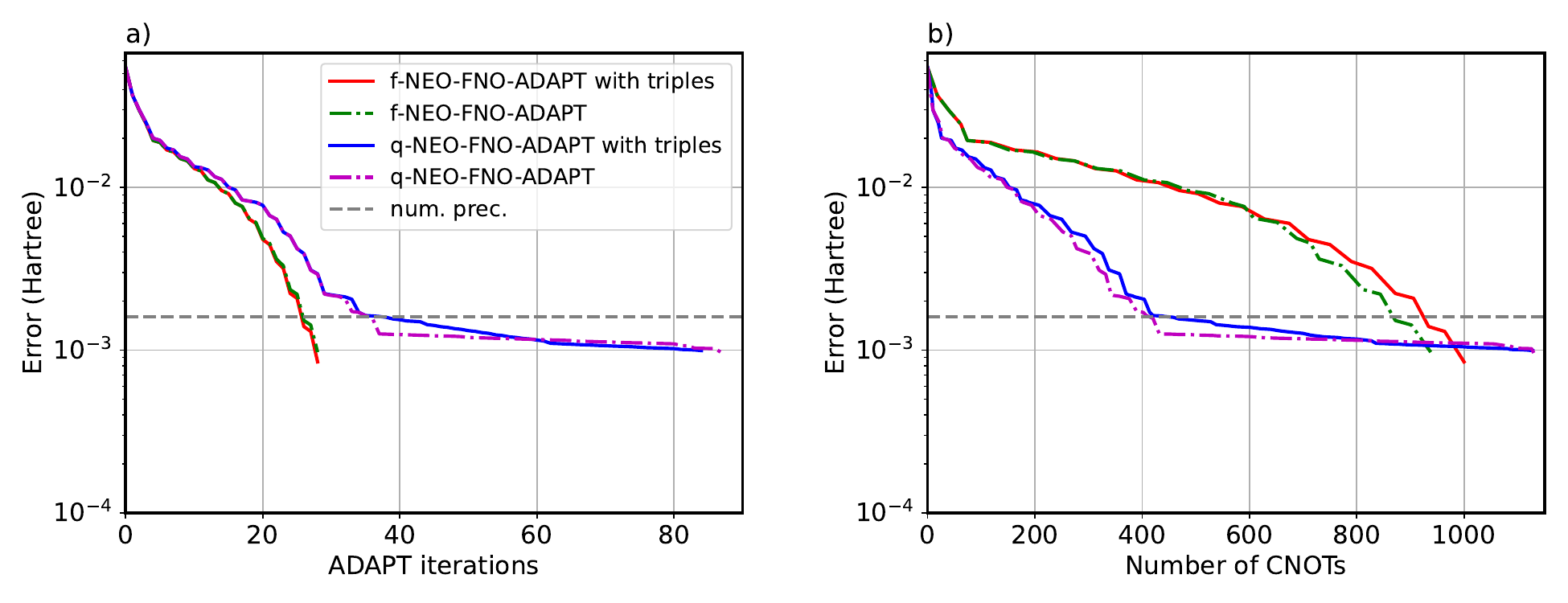}
  \caption{NEO-FNO-ADAPT simulations with fermionic and qubit operator pools for the H$_2$ molecule. Panel a) shows the error relative to the NEO-FNO-FCI energy as a function of the ADAPT iteration. Panel b) shows the error relative to the NEO-FNO-FCI energy as a function of CNOT count for the constructed circuit. The dashed lines correspond to numerical precision of 1.6~mHartree with respect to the NEO-FNO-FCI energy.}
  \label{fig:H2-sim}
\end{figure*}

\begin{figure*}[ht!]
  \centering
  \includegraphics[width=6.5in]{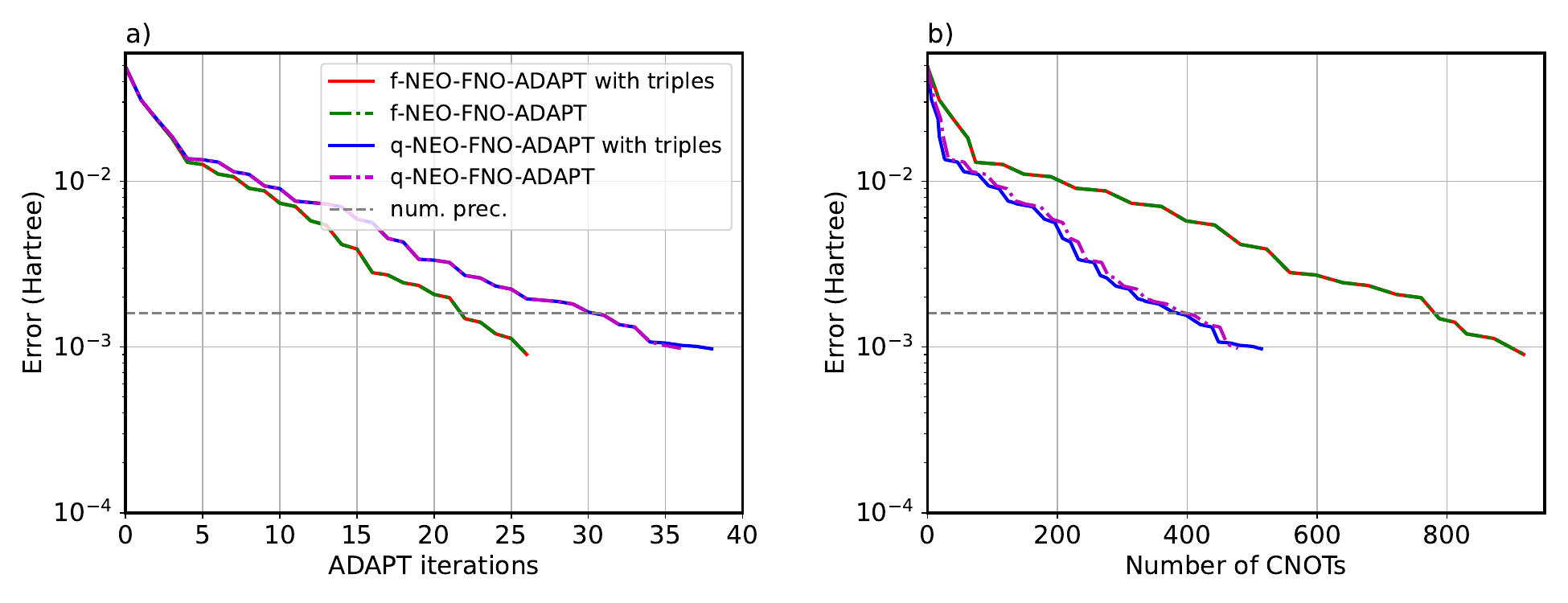}
  \caption{NEO-FNO-ADAPT simulations with fermionic and qubit operator pools for the D$_2$ molecule. Panel a) shows the error relative to the NEO-FNO-FCI energy as a function of the ADAPT iteration. Panel b) shows the error relative to the NEO-FNO-FCI energy as a function of CNOT count for the constructed circuit. The dashed lines correspond to numerical precision of 1.6~mHartree with respect to the NEO-FNO-FCI energy.}
  \label{fig:D2-sim}
\end{figure*}

The iterative convergence of the NEO-FNO-ADAPT-VQE simulations is illustrated in Fig.~\ref{fig:H2-sim} and~\ref{fig:D2-sim} for H$_2$ and D$_2$ molecules, respectively. Results are shown for multiple operator pools to study the convergence of different ADAPT setups in our framework. Solid lines indicate simulations with triple excitations (electron-electron-proton) present in the operator pool and dot-dashed lines without them. In both figures, panel a) shows the error as a function of the ADAPT iteration and panel b) shows the error as a function of the CNOT count of the constructed circuit. As is the case for the electronic ADAPT procedure, f-NEO-FNO-ADAPT converges to a numerical precision of 1.6~mHartree with respect to the NEO-FNO-FCI energy in fewer iterations than q-NEO-FNO-ADAPT, while resulting in a higher CNOT count for the constructed circuit. We find that for the simulation of the H$_2$ molecule, q-NEO-FNO-ADAPT exhibits plateaus in energy after reaching the numerical precision of 1.6~mHartree with respect to the NEO-FNO-FCI energy, slowing down further convergence. These plateaus are usually attributed to spin and number symmetry breaking caused by the q-ADAPT operators, but such occurrences were not detected in the present simulations. Furthermore, similar plateaus were not detected in the D$_2$ simulations. 
For comparison, in Figs.~S1 and S2 we show the corresponding NEO-ADAPT simulations in the NEO-HF basis, using the same number of orbitals as in the NEO-FNO case above.
The results show that these simulations converge in fewer iterations and require fewer CNOTs than their FNO counterparts, but, more importantly, recovering only a fraction of the correlation energy (see Table~S1) rendering them too inaccurate for meaningful quantum-chemical simulations.

Moreover, both figures unequivocally highlight that the energy difference between simulations with and without triple excitations in the operator pools are nearly identical, indicating that the inclusion of triple excitations has a minor effect on the energy in these cases. In particular, for the D$_2$ simulations, the f-NEO-FNO-ADAPT plots are exactly the same, which suggests that no triple excitations were selected throughout the ADAPT procedure. By contrast, in the q-NEO-FNO-ADAPT simulation, some triple excitations were chosen, as seen from a slight difference in CNOT counts in Fig.~\ref{fig:D2-sim}, although these additional operators seem to have a negligible effect on the convergence. 
In the case of H$_2$ simulations, both in f-NEO-FNO-ADAPT and in q-NEO-FNO-ADAPT simulations, a few triple excitations were chosen leading to a slightly more noticeable effect as illustrated in Fig.~\ref{fig:H2-sim}. In f-NEO-FNO-ADAPT, the CNOT count is therefore increased by the presence of triple excitations while the convergence rate remains substantially unaffected. A similar observation holds for the q-NEO-FNO-ADAPT simulation with increased difference in CNOT count after reaching our pre-defined numerical precision indicated by the dashed line in Fig.~\ref{fig:H2-sim}. These findings confirm that triple excitations do not contribute significantly beyond the precision scales considered here. We note that with an increase in the number of FNOs, the importance of triple and quadruple excitations will become more pronounced. This is evident from the fact that a relatively large energy difference between the NEO-CISD and NEO-FCI (see Table~\ref{table:table1}) originates predominantly from triple (electron-electron-proton and electron-proton-proton) and quadruple (electron-electron-proton-proton) excitations which are not considered in the NEO-CISD correlation approach. 

Furthermore, Table~\ref{table:table1}\ shows that the calculated energies for the NEO-FNO-UCCSD, f-NEO-FNO-ADAPT, and q-NEO-FNO-ADAPT methods are within numerical precision of 1.6~mHartree relative to the NEO-FNO-FCI data. Additionally, the CNOT count of the NEO-FNO-UCCSD simulations for both molecular systems is $\sim$5000. For comparison, an implementation of the NEO-UCCSD method where all orbitals are included in the active space would require roughly 70 million CNOT gates. The latter reveals that the FNO approximation itself reduces the gate count by four orders of magnitude in the case of these two systems. Finally, we note that for f-NEO-FNO-ADAPT, the CNOT count is $\sim$800\ which can be further reduced to $\sim$400 by resorting to q-NEO-FNO-ADAPT (see Table~\ref{table:table1}). The CNOT counts reported in Table~\ref{table:table1} correspond to quantum simulations without triple excitations.   

Lastly, we turn to a discussion of the energy difference between the H$_2$ and D$_2$ molecules summarized in Table~\ref{table:table1}. Because the NEO method inherently incorporates the zero-point energy in the energy calculations, it allows for an estimation of the kinetic isotope effect. Hence, by calculating the energy difference between the H$_2$ and D$_2$ molecules, we can get an estimate of the $\Delta$ZPE between different isotopes. The NEO-HF method predicts that the $\Delta$ZPE between the H$_2$ and D$_2$ is 13.8~kcal/mol. Experimentally, this value is determined to be 1.8~kcal/mol,~\cite{irikura2007experimental} which is very close to the value of 2~kcal/mol predicted with highly accurate and numerically nearly exact non-relativistic non-Born--Oppenheimer calculations~\cite{bubin2009non,bubin2010accurate}. The main source of the NEO-HF method's poor performance can be attributed to the lack of correlation effects between quantum particles due to the mean-field approximation~\cite{pavosevic2020chemrev}. Indeed, an inclusion of correlation effects through the singles and doubles excitations in the NEO-CISD and NEO-CCSD methods further systematically improves the estimate of $\Delta$ZPE. Finally, an inclusion of all three- and four-body interactions in the NEO-FCI method predicts the $\Delta$ZPE to be 3.9~kcal/mol. The remaining discrepancy of $\sim$15~mHartree between the NEO-FCI data and the numerically near exact prediction for $\Delta$ZPE~\cite{bubin2009non,bubin2010accurate}  originates thus predominantly from the use of a (albeit large) finite composite set of electronic and protonic basis functions as well as due to fact that the NEO Hamiltonian is not free from translational and rotational degrees of freedom~\cite{naka2005,hoshino2006elimination}. A further increase of the numerical bases would improve the agreement, albeit at a much higher cost. In passing, we note that the error due to the FNO approximation for the NEO-FNO-FCI method is only 0.3~kcal/mol. Notably, the NEO-FNO-CCSD, f-NEO-FNO-ADAPT as well as q-NEO-FNO-ADAPT data for $\Delta$ZPE not only cluster around 4.2~kcal/mol but are also in perfect agreement with the NEO-FNO-FCI data. In summary, all present findings corroborate our hypothesis that the developed quantum algorithms are capable of capturing a qualitatively correct description of the isotope effect.  

\section{Conclusion}

In this work, we have extended the hardware efficient ADAPT-VQE algorithm to the NEO framework. The proposed NEO-ADAPT-VQE approach allows for the calculation of important nuclear quantum effects, such as ZPE or nuclear delocalization, and is suitable for simulations beyond the Born--Oppenheimer approximation on currently available quantum devices. Furthermore, to make non-Born--Oppenheimer quantum chemical calculations beyond minimal basis sets routinely feasible, we introduce the FNO approximation for a compact and accurate truncation of the electronic and protonic unoccupied orbitals. We demonstrate with the example of H$_2$ and D$_2$ that the FNO approximation not only offers a robust way for reducing the computational cost within the NEO framework but that it simultaneously retains a major portion of the correlation energy. By contrast, only a fraction of the correlation energy is recovered if the truncation of orbital spaces is done on NEO-HF orbitals. Therefore, this approximation will also be central for the future development of other low-cost NEO wave function-based methods implemented on classical and quantum computers. Moreover, to put the NEO-FNO-ADAPT approach to a test, we investigate its performance by selecting operators for the ADAPT step from two distinct pools, that is, a fermionic pool (f-NEO-FNO-ADAPT) and a qubit pool (q-NEO-FNO-ADAPT). We demonstrate that either of the latter enable a CNOT count reduction by roughly an order of magnitude compared to NEO-FNO-UCCSD while maintaining numerical precision of 1.6~mHartree relative to the reference NEO-FNO-FCI method. In addition, by analyzing the NEO-FNO-UCCSD and NEO-UCCSD data, we estimate that the FNO approximation itself reduces the number of CNOT gates by four orders of magnitude due to lower  qubit requirements. Such extreme reductions of the CNOT gates are an essential step towards enabling accurate and chemically meaningful calculations within the NEO framework on near-term quantum devices. Work along these lines is in progress in our laboratories. 
Lastly, we illustrate that NEO-ADAPT-VQE is applicable to qualitatively correctly study the isotope effect for H$_2$ and D$_2$ by calculating the difference in ZPE. 

We expect that due to constant improvements of quantum hardware efficiency and the demonstrated accuracy of q-NEO-FNO-ADAPT, the latter will play a central role in future implementations and non-Born--Oppenheimer simulations on near-term quantum devices. In particular, it presents a promising approach for simulating the dynamics of tautomeric isomerization reactions introduced in Ref.~\citenum{Kovyrshin2023jpcl}. By employing q-NEO-FNO-ADAPT alongside variational time-evolution methods~\cite{yuan2019}, one can simulate these reactions in both adiabatic and nonadiabatic regimes using short, constant-size quantum circuits. Moreover, future developments will also include the extension of the developed methodology to larger molecular systems via the embedding techniques~\cite{rossmannek2023quantum} or development of approaches for calculation of excited states~\cite{ollitrault2020quantum,pavosevic2021neouccsd,pavosevic2023spinflip} to enable simulations of photoinduced nonadiabatic molecular processes. Overall, this work opens up a wide range of research directions for the treatment of nuclear quantum effects on classical and quantum computers.

\begin{acknowledgement}
The authors thank Dr. Christopher Malbon, Dr. Zehua Chen, and Prof. Yang Yang for helpful discussions. This work was supported by the Hartree National Centre for Digital Innovation, a collaboration between the Science and Technology Facilities Council and IBM. This research was also supported by the NCCR MARVEL, a National Centre of Competence in Research, funded by the Swiss National Science Foundation (grant number 205602) and the Wallenberg Center for Quantum Technology (WACQT). IBM, the IBM logo, and ibm.com are trademarks of International Business Machines Corp., registered in many jurisdictions worldwide. Other product and service names might be trademarks of IBM or other companies. The current list of IBM trademarks is available at \url{https://www.ibm.com/legal/copytrade}.

\end{acknowledgement}

\noindent\textbf{Supporting Information Available}: The supporting information includes: percent of the correlation energy recovered for different truncated NEO methods; NEO-ADAPT-VQE simulations for the H$_2$ and D$_2$ molecules.

\noindent\textbf{Conflict of interest}\\
The authors declare no conflict of interest.

\linespread{1}\selectfont
\bibliography{references}{}

\end{document}